# Immersive Commons: Designing Ephemeral Spatial Dialogue for Co-Located Mixed Reality Environments

Wuwu Ran[1], Mohammed Faraj Al-Suwaidi[2]

[1]University of York, [2]Qatar University

[1]wuwu.ran@york.ac.uk, [2]malsuwaidi@qu.edu.qa

**Abstract.** Head-mounted mixed reality (MR) often prioritises individual immersion, potentially drawing attention away from others sharing the room. This paper asks how MR can support co-located spatial dialogue through embodied interaction and activity visible to non-headset observers. We introduce Immersive Commons, a design framework, and Atmos Messages, a prototype combining gesture-based, ephemeral spatial messages with voice transcription, preset text, and a public floor-plan overview. Following a Research through Design approach, fifteen participants created 236 messages across two sessions, including 48 voice-authored messages. Questionnaire responses highlighted playfulness and social awareness, while observations suggested opportunities for exchange grounded in the architectural setting. Interaction clustered near the public display, although this was also where participants put on their headsets. These findings suggest the potential of ephemeral spatial messaging to support co-located exchange, with playfulness, shared visibility, and architectural context informing the design of more social and situated MR experiences.

## 1 Introduction

Head-mounted mixed reality (MR) layers responsive digital content onto the physical world, often centring the experience on an individual wearer. In shared settings, immersion must be considered alongside environmental awareness and social interaction. Research on VR exergames found that passthrough improved environmental awareness while reducing immersion, with spatial context also influencing social acceptability (Guo et al., 2024). This raises a further design question: how can MR support communication among people sharing an architectural interior?

Early mixed-reality research anticipated this concern: mixed-reality boundaries could keep differently equipped participants mutually aware across joined physical and synthetic spaces (Benford et al., 1998). More recent work has addressed the practical challenges of co-located headset use (McGill et al., 2020), while other studies have explored ways of making immersive experiences visible to others, including making immersive content accessible to non-headset participants through projection and shared displays (Gugenheimer et al., 2017; Jansen et al., 2020) and rendering the scene from a third-person perspective on a public display (Do et al., 2024). These approaches form part of a broader design space for shared immersive experiences reviewed by Bréhault et al. (2025). Yet technical co-presence does not itself produce dialogue. A system may synchronise objects while offering no socially meaningful reason to approach, read, or respond to them.

Embodied interaction locates meaning in practical action rather than screen content alone (Dourish, 2001). Human-Building Interaction treats digital systems and architecture as a coupled socio-technical arrangement (Alavi et al., 2019), while Harrison and Dourish (1996) distinguish abstract space from socially produced place. McCullough's (2013) ambient commons similarly frames attention as negotiated within everyday environments. Together, these perspectives inform an approach to MR communication in which bodily movement, spatial relationships, and the architectural setting are integral to how digital messages are encountered and interpreted.

Previous work by the authors has explored immersive technologies as active instruments within architectural education through a pedagogical framework that positions VR as a primary design tool within a fully virtualised architectural design studio (Al-Suwaidi et al., 2023). Subsequent work examined a multidisciplinary collaborative workflow in which immersive 3D sketching supported shared design development across architectural disciplines (Al-Suwaidi et al., 2024). Related investigations have also used gamified VR to explore users' agency within simulations of urban-screen surveillance (Ran, 2024). The present study introduces Immersive Commons as a design framework for communication through temporary digital traces in a shared physical setting. It treats the architectural interior itself as a medium through which co-located participants can communicate, leave traces, and respond to one another. Drawing on Stavrides (2016), the term commons foregrounds the practices through which people encounter, contribute to, and negotiate shared space. It identifies a design ambition whose conditions of participation must be examined in use.

This paper asks: How can MR support co-located spatial dialogue rather than personal visual consumption, and what roles do temporality, playfulness, public visualisation, and architecture play in that shift? The contribution is threefold. Conceptually, it reframes MR communication as spatial dialogue shaped by embodiment, temporality, and architectural context. Methodologically, it offers an evaluation framework organised around social legibility, co-located awareness, and spatial engagement for investigating collective interaction in shared interiors. Empirically, it presents a working multi-user co-located MR system combining first-person interaction with a third-person public display, together with findings from a two-session deployment that provide an initial assessment of the proposed framework's potential and limitations in practice.

## 2 Methodology

### 2.1 Research through Design

The study follows Research through Design (RtD), using a functioning artefact as the primary mode of inquiry (Zimmerman et al., 2007). Three overlapping phases structured the work: interaction prototyping; integration of networking, temporality, coordinate alignment, and logging; and situated deployment (Figure 1). Each phase produced both an artefact revision and an analytic question. Prototyping asked whether message-making could become a visible bodily act; integration asked which technical conditions were required for shared place; deployment asked whether the resulting traces supported awareness, encounter, and reply. A fault identified in the first field session was revised before the second, making the two sessions an RtD iteration pair rather than identical replications.

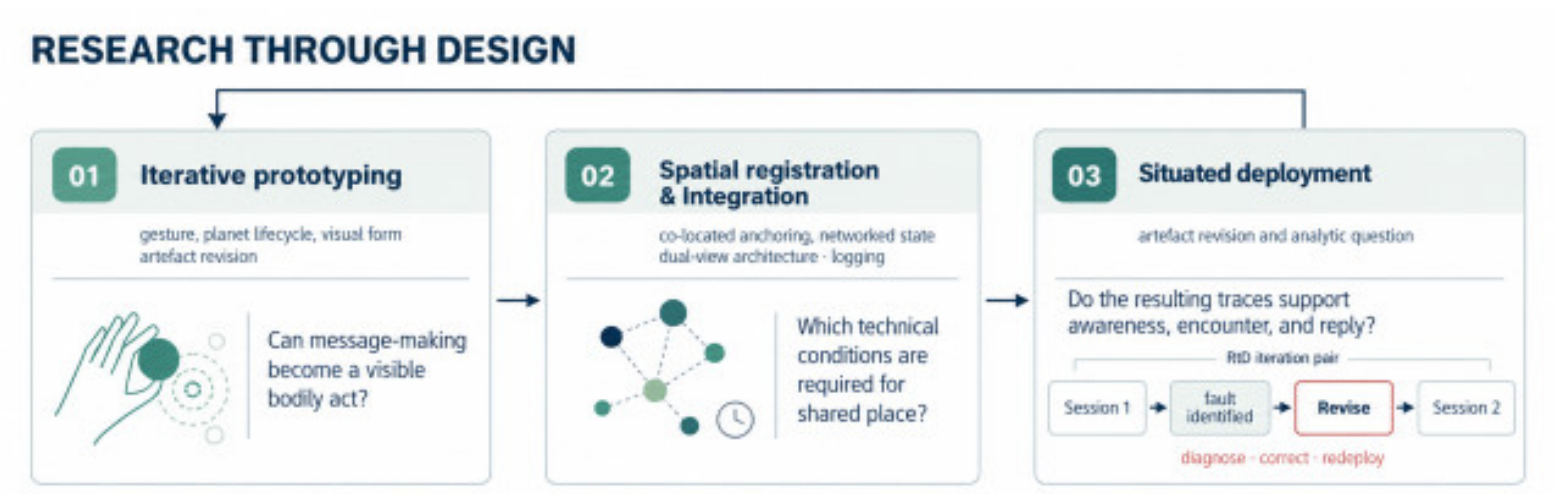

**Figure 1.** Three-phase Research through Design methodology. Source: Author, 2026.

### 2.2 Evaluation Framework

Evaluation was organised around three dimensions (Figure 2). Social legibility concerned whether communicative acts and contributions were intelligible to others, examined through questionnaire ratings and open-ended responses alongside observations of interaction and the public overview. Co-located awareness concerned participants' sense of others' presence and activity, examined through social-awareness ratings and written accounts of encountering and responding to messages. Spatial engagement concerned how participants used and interpreted the architectural setting, examined through floor-plan mapping of venue-registered interaction events and open-ended responses about message placement. Questionnaire ratings (Likert 1–5), written responses, and timestamped behavioural records were linked by participant ID and headset-wear windows and interpreted descriptively. Spatial analysis used interaction-event positions because head-pose logging was incomplete.

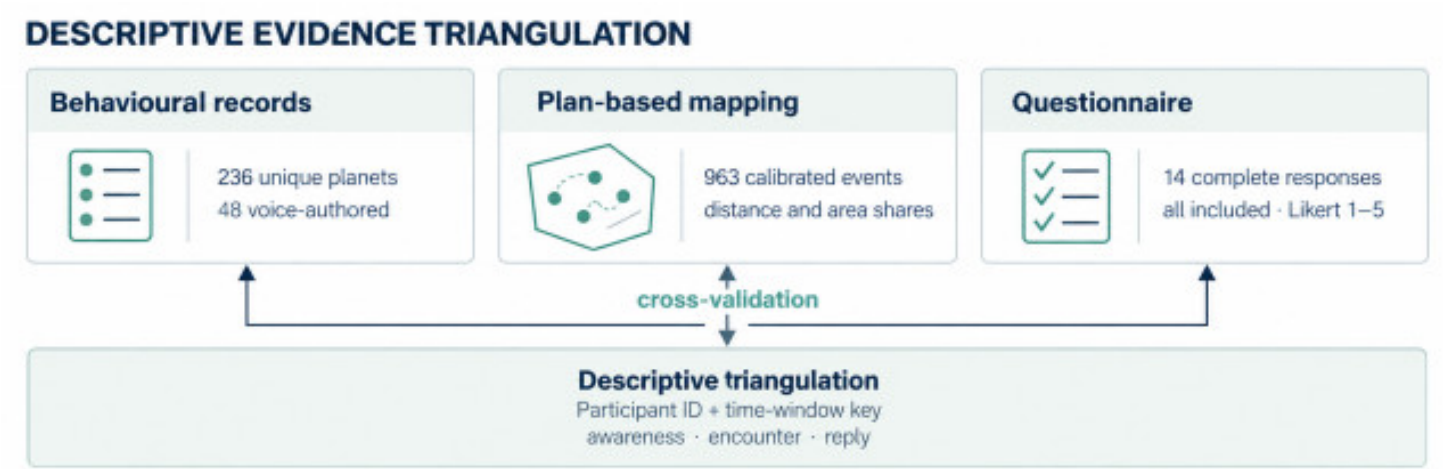

**Figure 2.** Descriptive triangulation of behavioural, spatial, and questionnaire data. Source: Author, 2026.

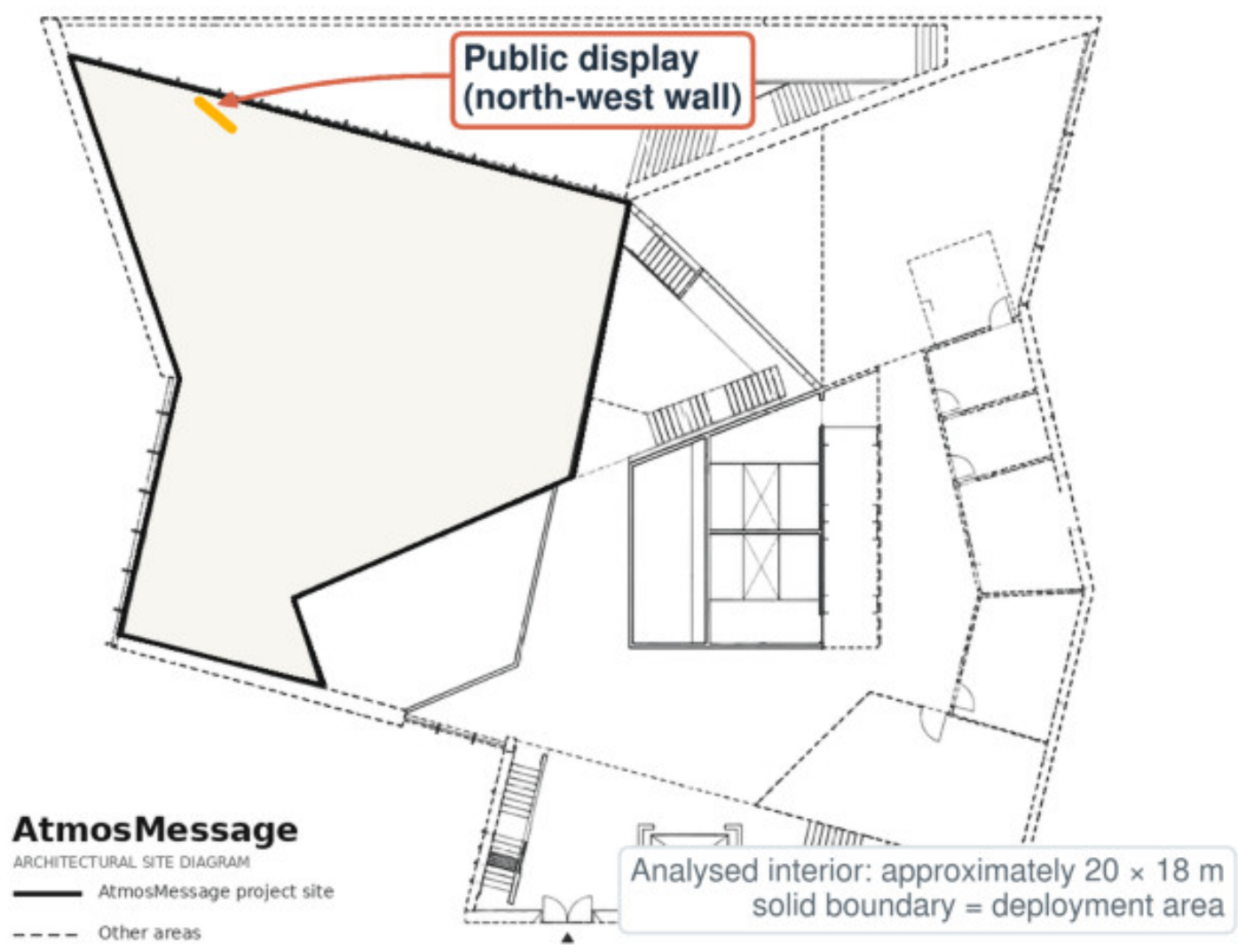

**Figure 3.** Study site: floor plan of a mixed-use academic interior with studio, exhibition, and circulation zones. Source: Author, 2026.

### 2.3 Participants, Site and Procedure

The deployment comprised two afternoon sessions, two days apart, with fifteen participants (P001–P015, including one research-team member). Of the fourteen respondents, eleven were aged 25–34 and nine reported no or limited prior MR experience. The study occupied an approximately 20 × 18 m mixed-use academic interior with studio, exhibition, and circulation zones (Figure 3). The site is representative of the semi-public interiors in which co-located MR is increas-ingly likely to be used; its physical requirements are addressed in the limitations. Two co-located headsets were worn in rotation while the overview ran continuously. Consent used a separate, named form; the anonymous questionnaire links to the behavioural record only through an assigned ID, and a facilitator's log of per-wear windows joins logs, questionnaires, and spatial analysis (Figure 4, Table 1). Fourteen questionnaires were collected; P005 withdrew (the headset could not be worn comfortably without glasses).

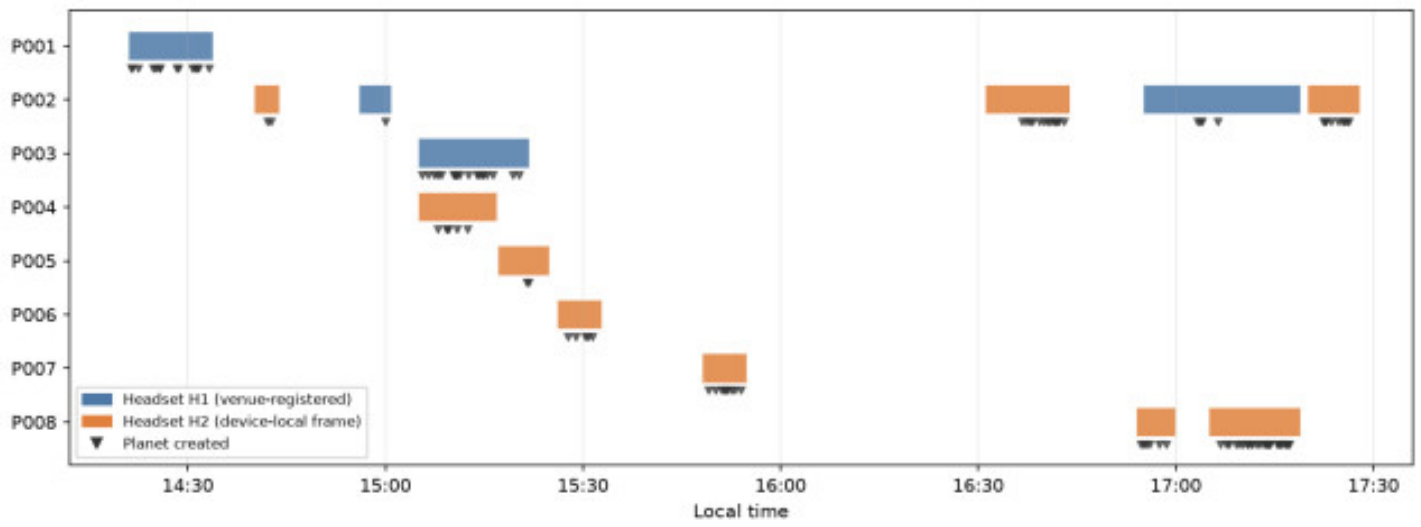

**Figure 4.** Session 1 timeline. Source: Author, 2026.

| Measure | Session 1 | Session 2 |
|---|---|---|
| Participants | 8 | 7 |
| Questionnaires | 7 | 7 |
| Message Planets | 125 | 111 |
| Voice-authored messages | 30 | 18 |
| Lifespan range | 10 min – 2 h (M ≈ 64 min) | 10 min – 2 h (M ≈ 73 min) |
| Venue-calibrated interaction events | 446 | 517 |
| Naturally co-present headset pairs | 1 | 2 |

**Table 1.** Deployment and dataset summary

## 3 System Design and Implementation

Atmos Messages is implemented in Unity 6 (URP) for Meta Quest 3 passthrough headsets, using Photon Fusion 2 in Shared Mode for networked multi-user state.

### 3.1 Interaction: Composing and Encountering Message Planets

The core interaction is a single continuous gesture: pinch, pull, release. This composes a Message Planet, a small semi-transparent sphere anchored at the point of release in shared space, and it reads as a physical act of drawing something out of the air rather than a menu operation. The gesture works with bare-hand tracking or a controller (pinch detection uses a hysteresis threshold with a tracking-loss grace period). Once anchored, a planet can be re-grabbed by any participant, transferring interaction authority while preserving its remaining lifetime; an encountered message thus becomes the occasion for a further communicative act (Discover-and-Dialogue).

Temporality is treated as an expressive parameter. Participants choose a lifespan from ten minutes to two hours using a rotary dial with live preview. The planet's initial size reflects the selected lifespan through a non-linear mapping, while shrinking near expiry signals its approaching disappearance. This treats ephemerality as a design material (Döring et al., 2013), with bounded lifetimes intended to lower the stakes of expression and focus attention on the present encounter (Bayer et al., 2016). Participants also choose one of eight visual types from a world-space ring.

Each planet carries a message for others to encounter. Messages can be composed through voice transcription or selected from presets. Speech recorded during composition is transcribed by a cloud speech-to-text service and displayed as text with the planet; presets provide a fallback when transcription is unavailable. User annotation of shared space has a long lineage, from GeoNotes' place-bound notes as a horizontal communication medium (Espinoza et al., 2001) to Blocks' collaborative, persistent AR structures (Guo et al., 2019). Within this lineage, Atmos Messages combines embodied composition, spatially anchored text, and deliberately bounded lifetimes to support communication through traces that others can discover, read, and respond to.

### 3.2 Co-located Spatial Registration

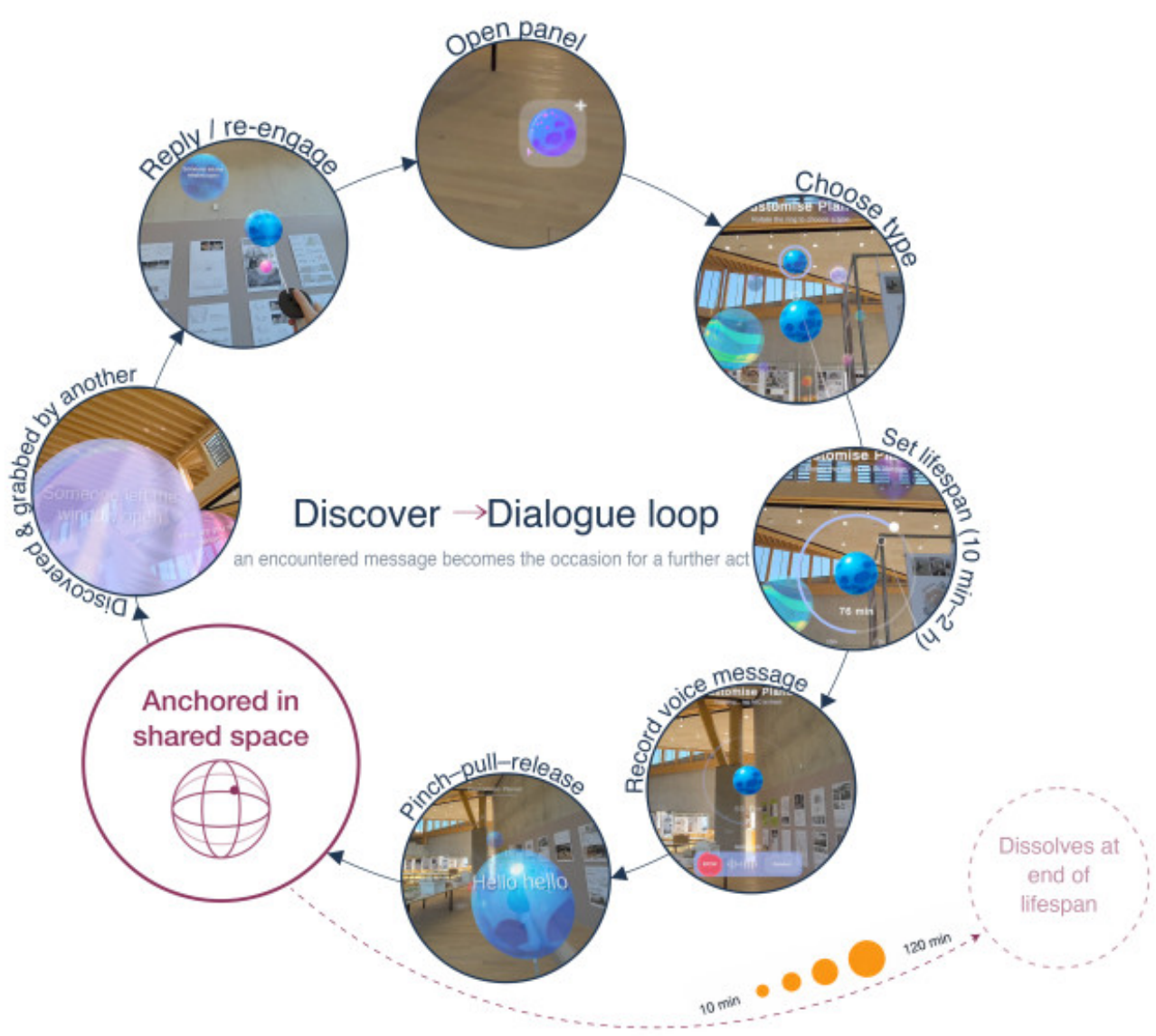

**Figure 5.** The Message Planet lifecycle in first-person view. Source: Author, 2026.

All participants must perceive each planet at the same physical location. The world origin is a fixed, physically marked point in the venue: at first launch the hosting headset creates a shared spatial anchor there, persisted and re-localised on every launch, so the origin and the overview's floor-plan cali-bration survive restarts and hand-offs. Guest headsets discover the co-location session automatically, requiring only a one-time device permission. Without co-located registration the session still runs in a device-local frame, and a reconnection watchdog rejoins the shared session after network interruptions. At each session's start, a lightweight two-point calibration (short-lived planets placed on the origin marker and at a surveyed corner) provides ground-truth correspondences from which the floor-plan projection's origin, rotation, and scale are solved exactly. These mechanisms proved consequential in deployment (4.2).

### 3.3 Dual First-Person / Third-Person Architecture

Alongside the first-person passthrough views, a public display presents a real-time, third-person, top-down overview of shared activity (Figure 6). Message Planets appear as coloured markers on the calibrated floor plan, with participants represented by live head markers. Drawing on approaches to making immersive content accessible beyond the headset (Gugenheimer et al., 2017; Jansen et al., 2020; Do et al., 2024), this scene-level representation was designed to help non-headset observers follow activity without relying on an individual wearer's changing viewpoint. The overview makes the distribution and movement of spatial messages publicly visible within the room.

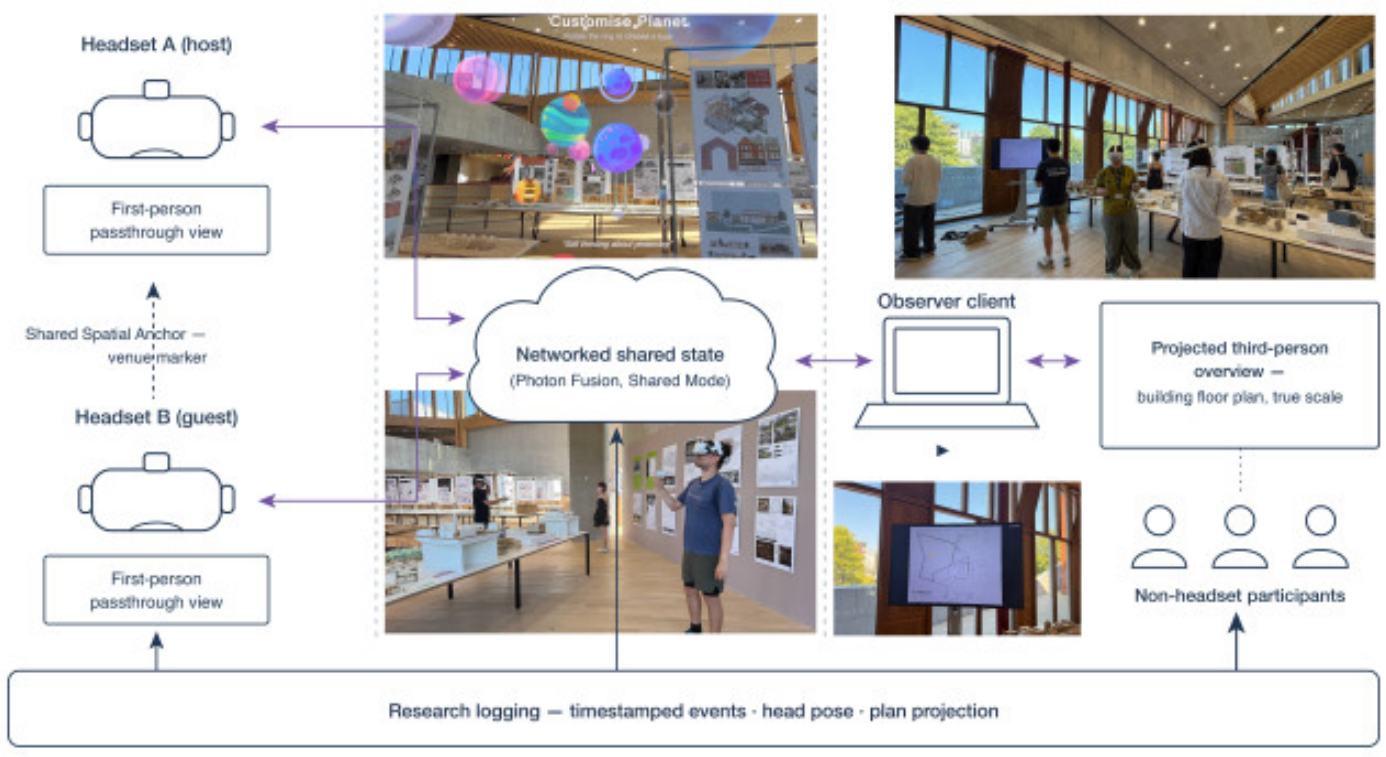

**Figure 6.** First- and third-person views in deployment. Source: Author, 2026.

### 3.4 Research Logging

Timestamped logs recorded creation, anchoring, release, re-grabbing, and spatial alignment. The system also recorded head-pose samples, but interruptions after reconnections prevented continuous trajectory analysis. Analysis therefore used timestamped interaction records and valid venue coordinates to produce event timelines, placement maps, and interaction-density maps.

## 4 Results

### 4.1 Technical Validation

Technical validation examined synchronisation, interaction behaviour, and spatial registration across clients. Synchronised instances, including a headset client and the observer, joined a shared session. Checks covered error-free builds, single-instance correctness, and cross-client consistency, assessed by comparing independently written event logs. Each planet's type and message text matched exactly across clients. Object motion followed a single-authority model, confirmed across clients and on the overview. In debug-accelerated sessions, every planet expired at its anchor time plus its selected lifespan, with re-grabbing transferring interaction authority without resetting the countdown. On-device validation confirmed 72 Hz rendering with 15 or more planets in view (98.7% of sampled seconds), centimetre-scale anchor re-localisation across restarts, and cross-headset registration to the same shared anchor.

### 4.2 Situated Deployment: Two Sessions

The system was deployed in two afternoon sessions: fifteen participants used two co-located headsets in rotation, the overview running continuously. In session one, a misconfigured spatial-sharing permission left one headset operating in a device-local coordinate frame. Networked messages remained available, but their positions were not consistently aligned with the physical venue across headsets. The permission was corrected before session two, when both headsets localised to the shared anchor. Session two therefore followed a technical revision. We distinguish message-level records from records with valid venue coordinates when reporting the two sessions.

Behavioural record. Participants created 236 unique Message Planets (125 and 111 per session). All eight types were used, with individual signatures: one participant placed 14 of 33 planets as a single favourite type, while another placed exactly one, at the maximum two-hour lifespan. Lifespans spanned the dial's range (10–120 min, mean ≈ 68). Voice authoring was substantial: 48 unique voice messages were transcribed onto planets, from greetings to situated observations of the building itself ("This building is really beautiful"; "Is this seat taken?"); the remainder used presets. Re-engagement, grabbing others' planets to read them, was frequent (65 grab events in session one alone).

Co-presence and dialogue. Three episodes of naturally co-present use occurred, with pairs wearing both headsets simultaneously, plus one facilitated cross-headset demonstration. In session two, the Discover-and-Dialogue loop operated unprompted: one participant reported that another "left his answer on the planet instead of saying it in real", a reply performed through the medium itself. Acquainted pairs used planets as directed asynchronous messages: one found "good luck with my ddl" and knew it was left for them. One participant created almost nothing yet participated intensively as a reader: authoring and listening are distinct roles within a spatial commons.

Spatial record. Figure 7a projects the venue-registered placements onto the floor plan at true scale (86 anchors; device-local session-one placements excluded, see Limitations); Figure 7b maps the density of all 963 venue-registered interaction events. Activity concentrated in the studio zone's exhibition-dense core, along the central model tables, and at thresholds. Interaction events clustered near the public display: 41% occurred within 3 m and 56% within 5 m, compared with 8% and 22% of the mapped site area, respectively. The concentration within 3 m was therefore approximately five times that expected under uniform spatial use. However, the display was located at the session starting point, where participants put on their headsets, so the observed clustering may also reflect headset fitting and initial use. Placement was also relational: one participant hid a planet beneath an exhibited model, "hiding a little secret for someone else to discover."

Questionnaire results are summarised in Figure 8. All fourteen questionnaires were analysed (Likert 1–5, descriptive), including the research-team member's, whose exclusion does not alter the pattern. Playfulness received the highest dimension mean (4.59; "felt playful" the highest item at 4.79), followed by overall evaluation (4.30), social awareness (4.19; "Reading messages left by others encouraged social engagement" at 4.36), message authoring (4.14), reduced-isolation items (4.11), and spatial engagement (4.05). Participants also associated playfulness with social exchange; one suggested that it "could act as a way to lower barriers of awkwardness". The weakest results are equally informative: social legibility was the lowest social dimension (3.76; "I could distinguish different participants' contributions" at 3.71), the lifespan dial's meaningfulness 3.50, and physical comfort weakest overall (3.46, headset weight cited).

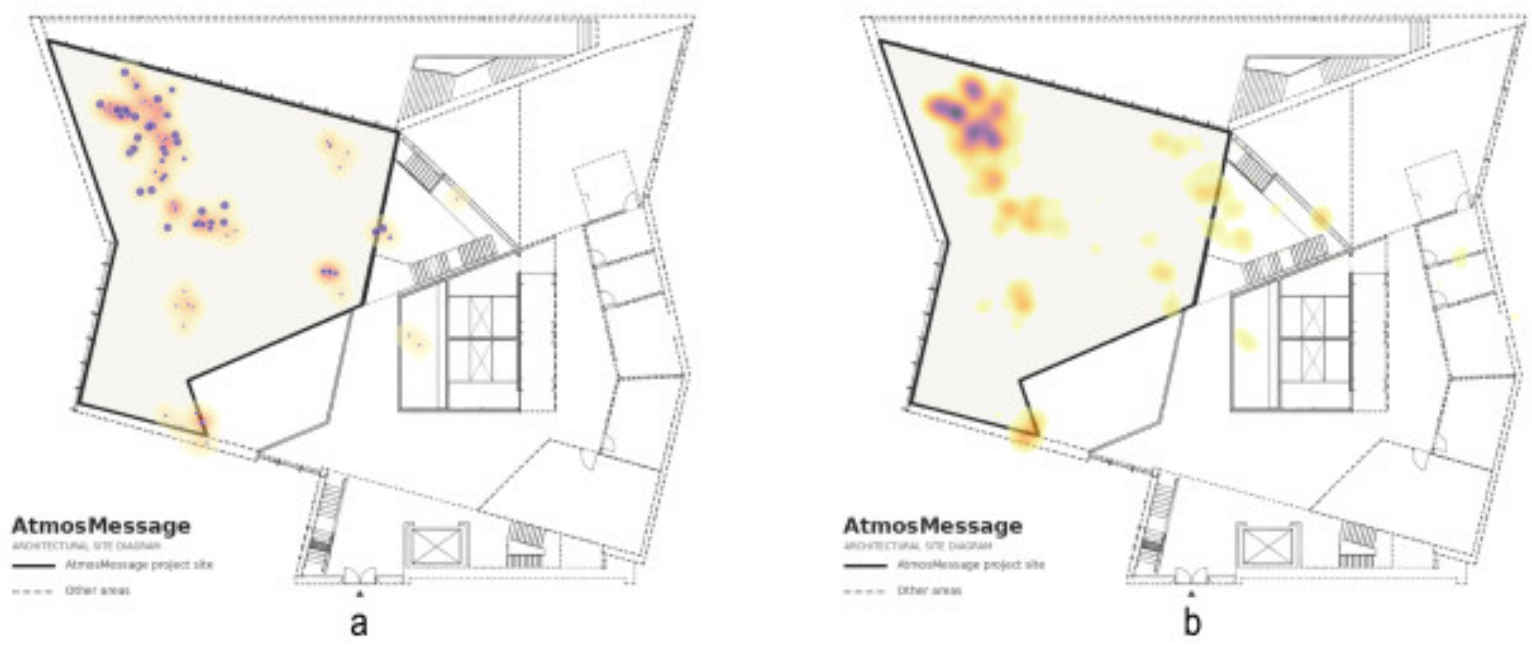

**Figure 7.** Venue-registered data across both sessions: (a) 86 Message Planet anchors mapped to scale; (b) density of 963 interaction events.

Triangulation and robustness. The data streams cohere at the individual level: the all-voice and preset-only creators rated voice authoring accordingly, and the participant who experienced the registration fault named it, un-prompted, as the improvement that mattered most (5.2). The network uplink dropped repeatedly; the reconnection watchdog restored the shared session within seconds each time. These incidents, legible in the logs, are themselves

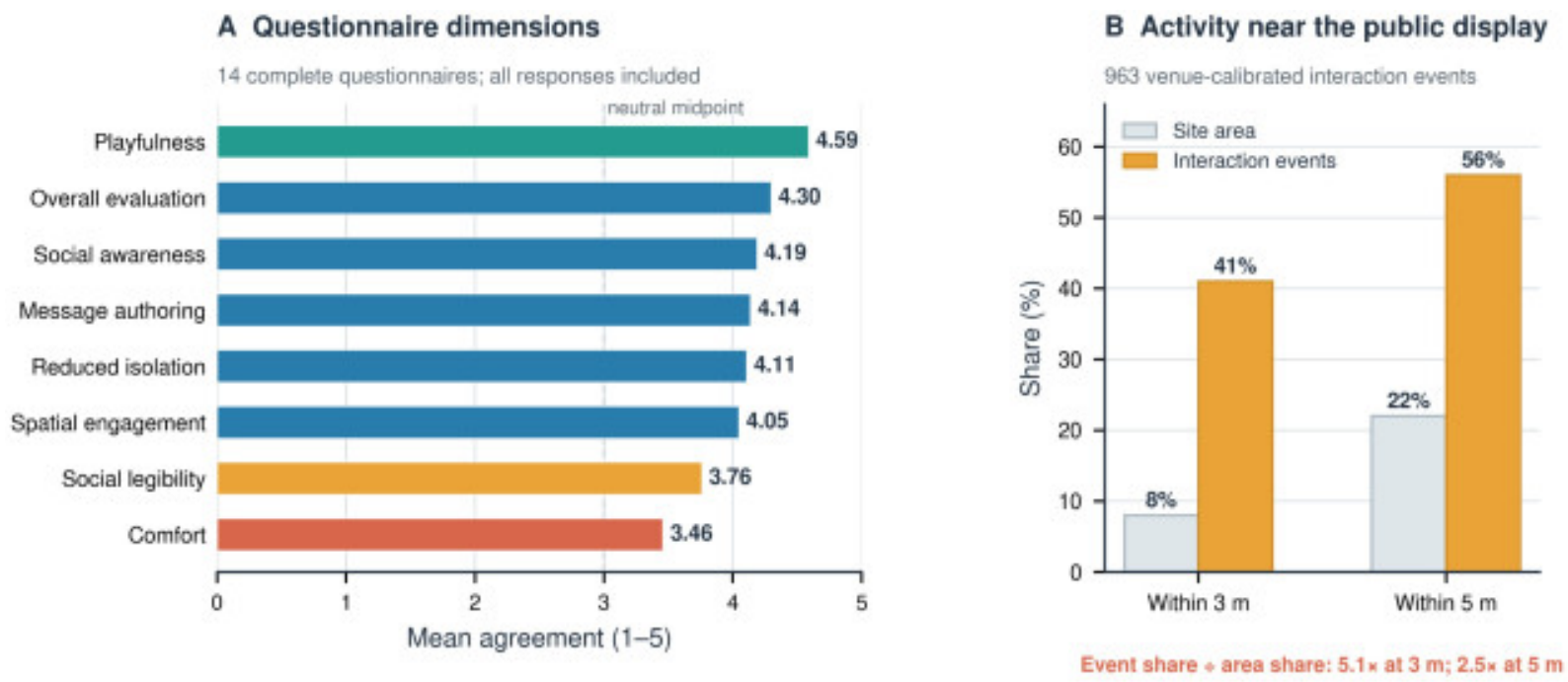

**Figure 8.** Questionnaire means and interaction clustering near the public display.
Source: Authors, 2026.

RtD findings: the emergent checklist (venue marker ritual, per-device permission, two-point calibration, dedicated network) is part of the framework's contribution.

## 5 Discussion

### 5.1 Immersive Commons as a Framework

Immersive Commons frames co-located MR communication as a spatial, embodied, and temporally bounded practice in which architecture is an active communicative medium. A planet hidden beneath an exhibited model (Figure 7a), and messages about the building or a seat, illustrate how communication drew meaning from its surroundings. Participants watched, imitated, and taught one another during composition, while lifespan choices spanned the available range. The exchanges reported in Section 4.2 further illustrate the communicative role of the shared interior: an unprompted reply was left on a Message Planet, and another message was recognised by its intended recipient. These examples suggest how spatially situated messages could connect participants through encounters in a shared place, with interpretation also shaped by existing social relationships. Gibson's account of perception and action helps interpret these placements as situated possibilities for engagement (Gibson, 2015), while Pallasmaa's emphasis on bodily and multisensory architectural experience directs attention to wearing a headset, moving, and orienting within the room (Pallasmaa, 2024). These connections remain interpretive: this study did not directly measure multisensory perception or isolate the effects of architectural form. The observations nevertheless ground the framework in relationships among messages, bodies, and material settings.

### 5.2 What the Deployment Teaches the Framework

Four findings refine the framework beyond design intention.

Playfulness may support social exchange. Playfulness received the highest questionnaire ratings, and one participant suggested that it could lower barriers of awkwardness. These findings resonate with perspectives on play as a mode of engagement (Sicart, 2014), an opportunity for encounters in public space (Stevens, 2007), and a resource for participation in urban media art (Guljajeva & Canet Sola, 2024). The toy-like character of Message Planets may offer co-located participants a shared pretext for interaction. In this sense, playfulness may function as a social lubricant within the framework, offering a design proposition for further study.

The public display may act as a spatial attractor. Designed to make interaction legible to observers, it was also near a concentration of activity: 41% of interaction events occurred within 3 m, approximately five times the proportion expected under uniform spatial use. However, the display was located at the designated starting point for the sessions, where participants put on their headsets. This clustering may therefore reflect attention to the shared overview, activity associated with headset fitting and initial use, or both; the present deployment cannot isolate the display's contribution. Comparisons with the overview absent or relocated, while keeping the headset fitting location fixed, would help distinguish these explanations and clarify the display's role in spatial dialogue.

Legibility of acts is not legibility of authors. The item concerning recognition of different participants' contributions received a mean rating of 3.71/5, highlighting a question for the framework: understanding a communicative act and identifying its author are distinct aspects of social legibility. Atmos Messages deliberately anonymises authorship, protecting low-stakes participation but capping the depth of dialogue: one can answer a message, not a person. Acquainted pairs' directed exchanges show participants routing around the limit with contextual knowledge. Early annotation systems already treated the signature as a regulator between privacy and recognisability (Espinoza et al., 2001); co-located MR sharpens the trade-off, since the anonymous author may be in the same room, and makes the balance between anonymity and attributability a primary design axis.

Registration is social infrastructure. The session-one permission fault left one headset's planets displaced for everyone else. Participants experienced this as a communication failure, not a technical blemish: "the coordinates were different than other headsets" was offered, unprompted, as the improvement that mattered most, and a session-two participant likewise named "better coordinated geolocation" after residual offsets. In co-located MR, spatial registration is the ground on which spatial dialogue stands, as a shared language is for conversation: its failure modes are social failure modes, and its establishment (marker, anchor, permission, calibration, network) must be designed, checked, and maintained as part of the commons' infrastructure. Diagnosed from the logs, corrected, and verified in session two, the fault also exercised the RtD loop as intended: the deployment protocol of 3.2 is not an appendix to the framework but one of its contributions.

### 5.3 Limitations

The study involved fifteen participants, including one research-team member, across two sessions in one venue. Questionnaire findings (n = 14) are descriptive, and eleven respondents were aged 25–34, limiting generalisation across ages. Participants' attention and perceptual conditions were not characterised, so accessibility for people with ADHD or differing sensory needs remains unestablished. Novelty may have influenced responses, and longer-term engagement was not assessed. Device-local placements from one headset in session one were excluded from spatial analysis; message counts, content, and timing remained usable. Residual alignment offsets limited spatial precision, while interrupted head-pose logging restricted analysis to interaction-event positions. Deployment also depended on suitable tracking conditions, a public display, and reliable connectivity. Physical comfort received the lowest dimension mean (3.46/5), and one participant withdrew because of headset-wearing difficulties. These observations identify comfort and access as conditions of participation without attributing difficulties to any particular diagnosis.

### 5.4 Future Work

Future studies should include a wider age range, including older adults, and involve participants with varied attentional and perceptual needs, including people with ADHD. Evaluation should record actual wear duration, fit with corrective lenses, perceived weight and balance, fatigue, discomfort, and reasons for withdrawal alongside communication measures. Adjustable stimulus density, breaks, seated use where appropriate, and meaningful routes to contribute without a headset would allow participation conditions to be examined empirically. Repeated encounters and comparisons of display configurations would also help distinguish initial novelty from sustained use.

## 6 Conclusions

This paper introduces Immersive Commons as a framework for co-located spatial dialogue and Atmos Messages as its prototype: a gesture-based, ephemeral spatial messaging system supporting voice authoring and combining first-person interaction with a third-person public display. The study suggests that playfulness and architectural context can provide opportunities for social exchange. It also highlights public display placement as a spatial design consideration, distinguishes the legibility of contributions from the recognisability of their authors, and identifies reliable spatial registration as social infrastructure. By treating the shared architectural interior as an active medium for communication, Immersive Commons offers a design stance, an evaluation framework, and initial evidence to inform more social and situated MR experiences.